\documentclass[a4paper]{jpconf}
\usepackage{graphicx}
\usepackage{psfrag}
\newcommand{\taut}{\tilde{\tau}}
\newcommand{\ttil}{\tilde{t}}
\begin{document}
\title{Quantum Dynamics of Electron-Nuclei Coupled System in a Double
Quantum Dot }
\author{\"Ozg\"ur \c{C}ak\i r and Toshihide Takagahara}
\address{Department of Electronics and Information Science,
Kyoto Institute of Technology, Matsugasaki, Kyoto 606-8585 JAPAN}
\address{CREST, Japan Science and Technology Agency, 4-1-8 Honcho,
Kawaguchi, Saitama 332-0012, JAPAN} \ead{cakir@kit.ac.jp}


\begin{abstract}
Hyperfine interaction of electron spins with nuclear spins, in
coupled double quantum dots is studied. Results of successive
electron spin measurements exhibit bunching due to correlations
induced via the nuclear spins. Further nuclear spins can be purified
via conditional electron spin measurements which lead to electron
spin revivals in the conditional probabilities. The electron spin
coherence time can be extended via conditional measurements. The
results are extended to a single electron on a single QD.
\end{abstract}

 Electron spins in semiconductor quantum dots are promising
candidates as building blocks for quantum information
processing\cite{Loss98,Taylor05,Imamoglu99}  due to their long
coherence times\cite{Golovach04,Semenov04}. The dominating
decoherence mechanism of electron spins is the hyperfine(HF)
interaction with the spins of the host
nuclei\cite{Coish05,Schulten78,Paget77, Merkulov02,Khaetskii03}.
Coherent manipulation of two-electron spin states has been achieved
in double quantum dots(QD) in recent experiments\cite{Petta05}, and
a detailed study of various aspects of hyperfine(HF) interaction and
electron spin decoherence became possible\cite{Johnson05}.

In section-\ref{sec_dqd}, we are going to introduce electron nuclei
coupled system in an electrically gated double QD occupied by two
electrons, then discuss bunching and revival in the results of
electron spin measurements. In the section-\ref{sec_sqd} the results
will be briefly extended to a single QD occupied by a single
electron.

\section{Electron nuclei coupled system in a coupled double QD \label{sec_dqd}}
 We consider an electrically gated double quantum dot(QD)
occupied by two electrons. Under a high magnetic field, s.t. the
electron Zeeman splitting is much greater than the hyperfine fields
and the exchange energy, dynamics takes place in the spin singlet
ground state $|S\rangle$ and triplet state of zero magnetic quantum
number $|T\rangle$,
\begin{eqnarray} H_e=JS_z+ r\delta h_z S_x,
\label{eq_hf} \end{eqnarray} where ${\bf S}$ is the pseudospin
operator with  $|T\rangle$ and $|S\rangle$ forming the $S_z$ basis.
 $\delta h_z=h_{1z}-h_{2z}$, where
$h_{1z}$ and $h_{2z}$ are the components of nuclear HF field along
the external magnetic field in the first and second dot,
respectively\cite{Coish05,Merkulov02}.
$r=t/\sqrt{t^2+(\delta/2+\sqrt{\delta^2/4+t^2})^2}$ is the amplitude
of the hyperfine coupling, which is determined by the gate voltages.
$\delta$ is the detuning which is a linear function of gate voltage
differences, and $t$ is the tunneling coefficient. When $\delta\gg
t$, the ground state singlet state corresponds to the case where
both electrons are localized in the same dot and HF coupling is
switched off, $r\rightarrow 0$. The opposite limit $\delta\ll -t$
corresponds to the singlet state where electrons are located in
different dots, and HF coupling is maximized $r\rightarrow 1$.

\subsection{Bunching in electron spin measurements}
Now we show that by electron spin measurements the coherent behavior
of nuclear spins can be demonstrated.
 Electron spins are initialized in the singlet state and the nuclear spin states are initially in a mixture of  $\delta h_z$ eigenstates,
$\rho(t=0)=\sum_n p_n\rho_n|S\rangle\langle S|$, where $\rho_n$ is a
nuclear state with an eigenvalue of $\delta h_z=h_n$ and satisfies
$Tr(\rho_n)=1$. $p_n$ is the probability of the hyperfine field
$\delta h_z$ having the value $h_n$.

In the unbiased regime $\delta\ll -t$, the nuclear spins and the
electron spins interact for a time span of $\tau$. Then the gate
voltage is swept adiabatically to a high value(s.t. $\delta\gg t$),
in a time scale much shorter than HF interaction time, leading to
the state, $ \rho=\sum_n p_n\rho_n|\Psi_n\rangle\langle \Psi_n|$,
 where $|\Psi_n\rangle=\alpha_n|S\rangle+\beta_n|T\rangle$, with
 $
 \alpha_n=\cos(\Omega_n \tau/2)+iJ/\Omega_n \sin(\Omega_n \tau/ 2)$,
 $\beta_n=
 -ih_n/\Omega_n\sin(\Omega_n \tau/2)\label{ampl}
 $
 and  $\Omega_n=\sqrt{J^2+h_n^2}$ is the Rabi frequency.

Next a charge state measurement is performed which detects a singlet
or triplet state\cite{Petta05}. Probability to detect the singlet
state is $\sum_n p_n|\alpha_n|^2$, and the triplet state is $\sum_n
p_n|\beta_n|^2$. Subsequently one can again initialize the system in
the singlet state of electron spins, and  turn on the hyperfine
interaction for a time span of $\tau$, and perform a second
measurement. In general over $N$ measurements, the probability of
$k$ times singlet outcomes is
 \begin{eqnarray}
 P_{N,k}=\bigl(^N_{\,k}\bigr)\langle
|\alpha|^{2k}|\beta|^{2(N-k)}\rangle.\label{Pqm}
 \end{eqnarray}
where $\langle\ldots\rangle$ is the ensemble averaging over the
hyperfine field $h_n$\cite{Merkulov02}.
 Here the key assumption is
that nuclear states preserve their coherence over $N$ measurements,
thus the measurements are not independent due to nuclear memory. One
can easily contrast this result with the semiclassical(SC) result
for which nuclear spins are assumed to be purely classical, whereas
electron spins are taken to obey quantum mechanics\cite{Merkulov02}.
In SC case results of successive measurements
 are independent and the probability for obtaining $k$ times singlet results over $N$ measurements is given by,
\begin{eqnarray}
 P'_{N,k}=\bigl(^N_{\,k}\bigr)\langle|\alpha|^2\rangle^{k}\langle|\beta|^2\rangle^{(N-k)}.\label{Psc}
\end{eqnarray}

In the SC case  the probability distribution (\ref{Psc}) obeys
simply a Gaussian distribution with mean
$k=N\langle|\alpha|^2\rangle$, and variance
$N\langle|\alpha|^2\rangle\langle|\beta|^2\rangle$, as
$N\rightarrow\infty$. However, in quantum mechanical(QM) treatment
of nuclear spins, the probability distribution (\ref{Pqm}) may
exhibit different statistics depending on the initial nuclear state.
If the SC distribution of $h_n$ is characterized by the same
distribution as in QM case, the two probability distributions
(\ref{Pqm}) and (\ref{Psc}) yield the same mean value, $
\overline{k}=N\langle|\alpha|^2\rangle$, however with distinct
higher order moments. If the distribution of initial nuclear state
$p_n$ has a width $\Delta$, then for HF interaction time $\tau\geq
1/\Delta$, the SC and QM distributions start to deviate from each
other. They yield the same distribution only when the initial
nuclear state is in a well defined eigenstate of $\delta h_z$, i.e.
when $\Delta=0$.

\begin{figure}[h!]
\begin{minipage}{18pc}
\includegraphics[width=18pc]{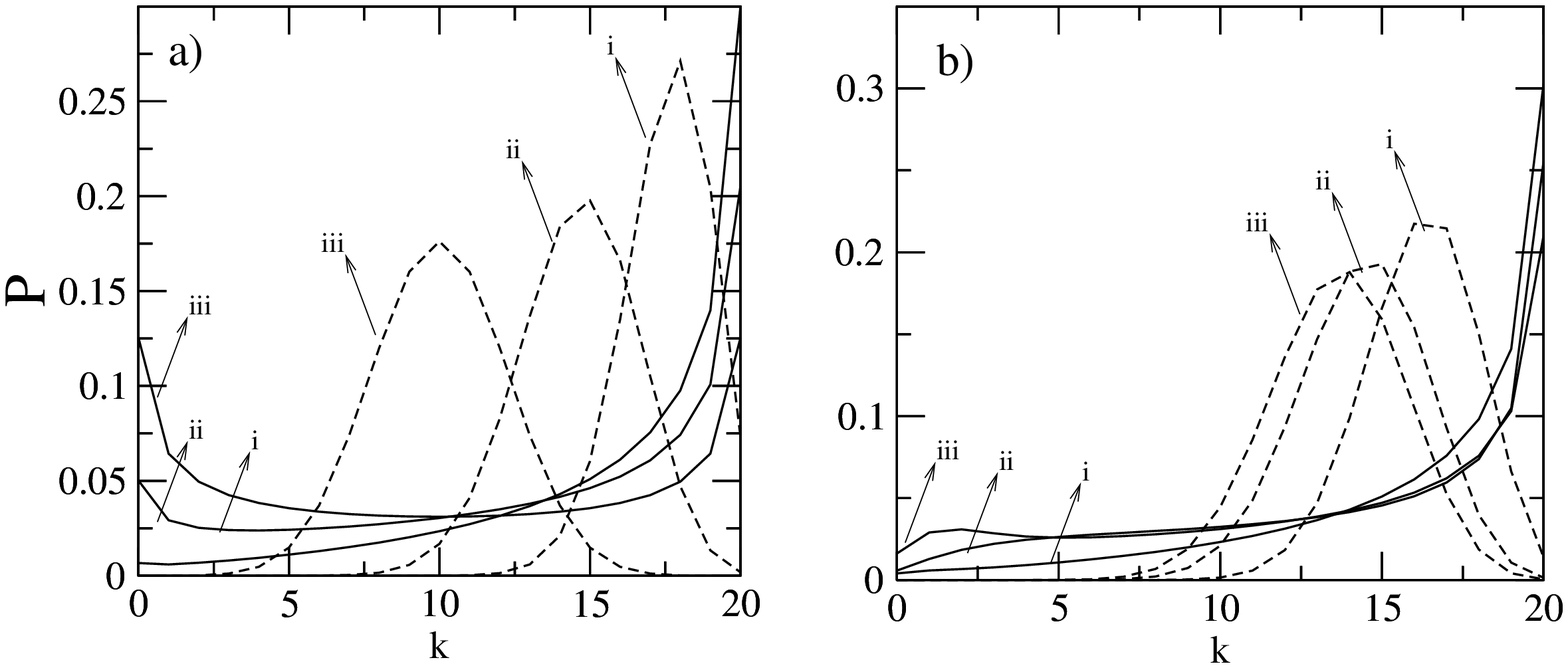}
\caption{Double QD: Probability distribution  at $N=20$ measurements
for $k={0,1,\ldots,20}$ times singlet detections, for QM(solid
lines), SC(dashed lines). Two cases of the exchange energy are
considered a) $J=0$ b) $J/\sigma=0.5$ for HF interaction times
$\sigma\tau=$ i)$0.5$, ii)$1.5$, iii)$\infty$. \label{Fig_20meas}}
\end{minipage}\hspace{2pc}%
\begin{minipage}{18pc}
\psfrag{a}{$k$} \psfrag{b}{$P$}
\includegraphics[width=18pc]{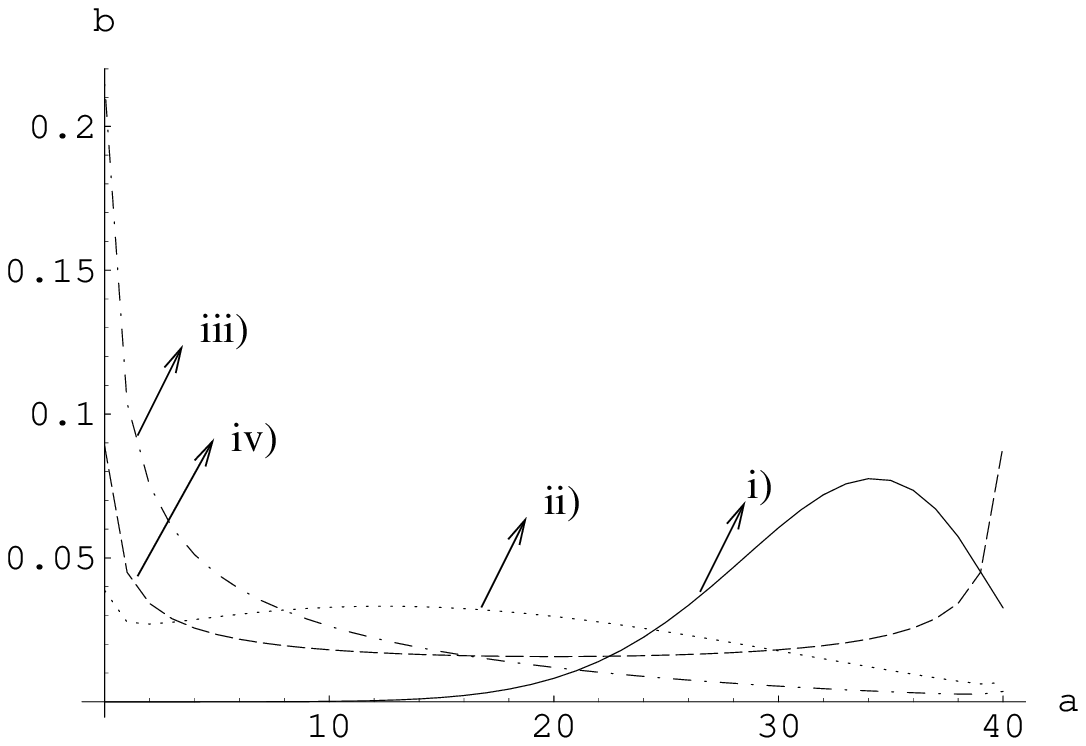}
\caption{Single QD: Probability distribution  at $N=40$ measurements
for $k={0,1,\ldots,20}$ times $|+\rangle$ detections at HF
interaction times  $\sigma\tau=$ i)$0.3$, ii)$0.6$, iii)$0.9$
iv)$\infty$. \label{figSQD}}
\end{minipage}
\end{figure}

In particular we are going to consider the case when the nuclear
spins are initially randomly oriented; probability distribution for
hyperfine fields obeying a Gaussian distribution $p_n\rightarrow
p[h]=1/\sqrt{2\pi\sigma^2}\exp[-h^2/2\sigma^2]$. In Fig.
\ref{Fig_20meas}, for $N=20$ measurements, $P_{N,k}$ is shown for HF
interaction times $\sigma\tau=0.5, 1.5, \infty$. For $\tau=0$, the
probability for both SC and QM cases is peaked at $k=20$. However,
immediately after the HF interaction is introduced, the probability
distributions show distinct behavior. The SC distribution converges
to a Gaussian distribution. In QM case the probabilities bunch at
$k=0,20$ for $J=0$, and when $J/\sigma=0.5$ those bunch at $k=20$
only. As $J$ is increased above some critical value, no bunching
takes place at $k=0$ singlet measurement.

\subsection{Electron spin revivals\label{sec_esr}}

\begin{figure}[h!]
\includegraphics[width=0.9\textwidth]{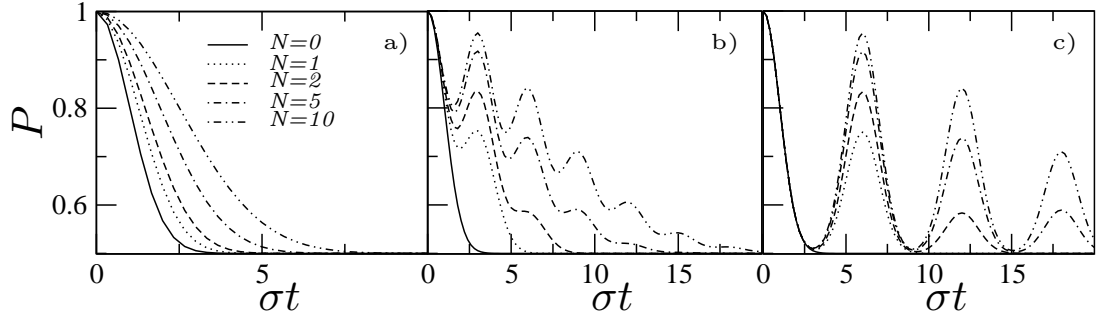}
\caption{Conditional probability for singlet state detection as a
function of HF interaction time $\sigma t$, subject to
$N=0,1,2,5,10$ times prior singlet state measurements and for HF
interaction times a)$\sigma\tau=1.0$, b)$\sigma\tau=3.0$,
c)$\sigma\tau=6.0$. \label{Fig-cond}}
\end{figure}

The modified nuclear spectrum leads to correlations between the
successive electron spin measurements. Depending on the results of
previous measurement, one may decrease the singlet-triplet mixing.
 As a particular example consider the
case: Starting from a random spin configuration, $N$ successive
electron spin measurements are performed, each following
initialization of electron spins in the spin singlet state and a HF
interaction of duration $\{\tau_i,i=1\ldots N\}$ and all outcomes
turn out to be singlet. Then again HF interaction is switched on for
a time $t$, and the $(N+1)$th measurement is carried out. The
conditional probability to detect the singlet state is given by
 \begin{eqnarray}
 P=\frac{\sum(^{~2}_{s_1})(^{~2}_{s_2})\ldots(^{~2}_{s_{N+1}})e^{-\frac{1}{2}\bigl[(s_1-1)\taut_1+(s_2-1)\taut_2+\ldots
+(s_N-1)\taut_N+(s_{N+1}-1)\ttil\bigr]^2}}{
4\sum(^{~2}_{s_1})(^{~2}_{s_2})\ldots(^{~2}_{s_N})e^{-\frac{1}{2}\bigl[(s_1-1)\taut_1+(s_2-1)\taut_2+\ldots
+(s_N-1)\taut_N\bigr]^2} },
 \label{condprob}
 \end{eqnarray}
 where the sums run over $s_i=0\ldots 2$ and $\taut_i=\sigma\tau_i$.
For the particular case $\tau_1=\tau_2=\ldots=\tau_N=\tau\gg
1/\sigma$,
 the initial state is revived at $t=n\tau, \,(n=1,2,\ldots,N)$ with
 a decreasing amplitude, $P\simeq1/2+\sum_{s=0}^{N}(
^{2N}_{~s})e^{\frac{-\sigma^2}{2}(t-(N-s)\tau)^2}/4(^{2N}_{~N})$. In
 Fig. \ref{Fig-cond} the conditional probabilities(\ref{condprob})
 are shown for $\sigma\tau=1.0, 3.0, 6.0$ subject to $N=0,1,2,5,10$
 prior singlet measurements in each. Revivals are observable only
 for $\sigma\tau>1$, because the modulation period of the nuclear
 state spectrum characterized by $1/\tau$ should be smaller than the
 variance $\sigma$.

From (\ref{condprob}), number of revivals can be increased with
various choices for the ratios of HF interaction times $\tau_i$.
 The underlying mechanism of revivals is purification of nuclear
spins by the electron spin measurements. The purity of a system
characterized by the density matrix $\hat{\rho}$ is given by ${\cal
P}=Tr\rho^2$. As an example we are again going to consider the
nuclear state prepared by $N$ successive electron spin measurements
with singlet outcomes, each followed by HF interaction times
$\tau_{1}\ldots \tau_N$,
\begin{eqnarray}
{\cal P}=\frac{1}{\cal D}\frac{\sum_{s_i=0}^
4(^{~4}_{s_1})(^{~4}_{s_2})\ldots(^{~4}_{s_N})e^{-\frac{1}{2}\bigl[(s_1-2)\taut_1+(s_2-2)\taut_2+\ldots
+(s_N-2)\taut_N\bigr]^2}}{\bigl[ \sum_{s_i=0}^
2(^{~2}_{s_1})(^{~2}_{s_2})\ldots(^{~2}_{s_N})e^{-\frac{1}{2}\bigl[(s_1-1)\taut_1+(s_2-1)\taut_2+\ldots
+(s_N-1)\taut_N\bigr]^2}\bigr]^2 }.\label{purity}
\end{eqnarray}
${\cal D}$ is the dimension of the Hilbert space for the nuclear spins. For a fixed ratio of $\tau_1:\tau_2:\ldots:\tau_N$, purity
(\ref{purity}) is a monotonically increasing function of time. For
$\sigma\tau_i\gg 1$, one can attain various asymptotic limits for
the purity.  For instance for $N=2$, there are three asymptotic
limits;when a)$\tau_1=2\tau_2$ then ${\cal P}=11/4{\cal D}$,
b)$\tau_1=\tau_2$ then ${\cal P}=35/18{\cal D}$, c)otherwise ${\cal P}=9/4{\cal D}$.
For $N=2$ with $\tau_1=2\tau_2=2\tau\gg 1/\sigma$,  the conditional
probability (\ref{condprob}) is given as follows,
\begin{eqnarray}
P\simeq\frac{1}{2}+\frac{1}{8}\bigl\{ e^{-\frac{(\ttil-3\taut)^2}{2} }+
2e^{-\frac{(\ttil-2\taut)^2}{2} }+ 3e^{-\frac{(\ttil-\taut)^2}{2} }+
4e^{-\frac{\ttil^2}{2} } \bigr\},\label{cond2}
\end{eqnarray}
whereas for $\tau_2=\tau_1=\tau\gg 1/\sigma$,
\begin{eqnarray}
P\simeq\frac{1}{2}+\frac{1}{12}\bigl\{ e^{-\frac{(\ttil-2\taut)^2}{2} }+
4e^{-\frac{(\ttil-\taut)^2}{2} }+6e^{-\frac{\ttil^2}{2} }
\bigr\}.\label{cond3}
\end{eqnarray}
As the purity of nuclear spins increase, more revivals are present
with an increased amplitude.

\section{Electron spin bunching and revivals in a single QD\label{sec_sqd}}
Now we are going
to consider a single electron on a single QD. Under external field
$B$, the system is governed by the Hamiltonian,
\begin{eqnarray}
H=g_e\mu_BB S_z +g_n\mu_n B\sum I^{(j)}_z  + {\bf h}\cdot{\bf
S}.\label{hfsqd}
\end{eqnarray}
In (\ref{hfsqd}), the first two terms are electron and nuclear
Zeeman energies respectively, and the last term is the HF
interaction where ${\bf h}$ is the HF field. When electron Zeeman
energy is much greater than rms value of HF fields, viz.
$g_e\mu_BB\gg\sqrt{\langle h^2\rangle} $, flip-flop terms are
suppressed and the Hamiltonian (\ref{hfsqd}) becomes, $H\simeq
g_e\mu_BB S_z + h_z S_z$. Up to a unitary rotation, with $B=0$, this
is equivalent to the Hamiltonian (\ref{eq_hf}) with $J=0$.
$|\pm\rangle=(|\uparrow\rangle)\pm|\downarrow\rangle/\sqrt{2}$ states
are coupled by HF interaction  with $|\uparrow(\downarrow)\rangle$
being the eigenstates of $S_z$. Each time the electron is prepared
in $|+\rangle$. Next it is loaded onto the QD, then removed from the
QD after some dwelling time $\tau$. Next spin measurement is
performed in $|\pm\rangle$ basis. Essentially the same predictions
as that of double QD can be made for this system, namely electron
spin bunching and revival.

In Fig. \ref{figSQD}, for $N=40$ measurements, the QM probability
distribution of $P_{N,k}$ is shown at electron Zeeman energy
$\epsilon=g_e\mu_B\hbar/2=3\sigma$, for $\sigma\tau=0.3,~,0.6,~0.9,~\infty$. It is
seen that contrary to the double QD, the population bunches at
$|-\rangle$ states at times $\tau\sim \pi/\epsilon$, but then relaxes to the
equilibrium distribution cf. Fig. \ref{Fig_20meas}.

Next we are going to consider electron spin revivals.  For instance
after $N$ times HF interaction of duration $\tau\gg 1/\sigma$, each
followed by $|+\rangle$ measurement, the conditional probability for
obtaining $|+\rangle$ in the $(N+1)$th step followed by a HF
interaction of duration $t$ is given as, $ P\simeq
1/2+\sum_{s=0}^{N}(
^{2N}_{~s})e^{-\sigma^2(t-(N-s)\tau)^2/2}\cos\epsilon[t-(N-s)\tau]/4(^{2N}_{~N})$.
 This is
essentially the same result for that of a double QD discussed in
section-\ref{sec_esr}.

\section{Discussion and conclusion}

The randomization of nuclear spins will lead to loss of memory
effects described above. The nuclear state conditioned on the
electron spin measurements will decohere during time interval
between the successive measurements, i.e., when the HF interaction
is switched off. Thus, the main decoherence mechanism of nuclear
spins is due to intrinsic nuclear dipole-dipole interactions. In
double quantum dots the duration of the cycle involving electron
spin initialization and measurement is about $10
\mu$s\cite{Petta05}. Since the nuclear spin coherence time
determined mostly by the nuclear spin diffusion is longer than about
several tens of ms\cite{Paget82,Giedke05,Paget77}, the bunching for
$N$ successive measurements up to $N>1000$ can be observed. The same
holds for the number of revivals that can be observed.

We have studied the quantum dynamics of the electron-nuclei coupled
system in QD's. The bunching of results of the electron spin
measurements and the revival in the conditional probabilities are
emerging features of coherence of nuclear spins. The underlying
mechanism is the correlations between successive measurements
induced via nuclear spins and the increase in the purity of the
nuclear spin state through the electron spin measurements. This
mechanism is expected to lead to the extension of the electron spin
coherence time.

\end{document}